\documentclass{IEEEtran}
\usepackage{amsmath,amsfonts}
\usepackage{color}
\usepackage{array}
\usepackage{algorithmic}
\usepackage[linesnumbered,ruled,vlined]{algorithm2e}
\usepackage[caption=false,font=normalsize,labelfont=sf,textfont=sf]{subfig}
\usepackage{textcomp}
\usepackage{stfloats}
\usepackage{url}
\usepackage{times}
\usepackage{verbatim}
\usepackage{graphicx}
\usepackage[nocompress]{cite}
\usepackage{longtable}
\usepackage{tabularx}
\usepackage[utf8]{inputenc}
\usepackage[T1]{fontenc}
\usepackage{lipsum}
\usepackage{multirow}
\usepackage{mathtools}
\usepackage{titlesec}
\usepackage[utf8]{inputenc}
\usepackage[linesnumbered,ruled,vlined]{algorithm2e}
\usepackage{amsmath, amssymb}
\makeatletter
\renewcommand\paragraph{\@startsection{paragraph}{4}{\z@}%
                                    {1.3ex plus 1.5ex minus 0.2ex}%
                                    {0.7ex plus .2ex}%
                                    {\normalfont\normalsize\itshape}}
\renewcommand\subparagraph{\@startsection{subparagraph}{5}{\z@}%
                                       {1.3ex plus 1.5ex minus 0.2ex}%
                                       {0.7ex plus .2ex}%
                                       {\normalfont\normalsize\itshape}}
\makeatother

\begin{document}

\linespread{0.998}
\title{Robust Precoding for Resilient Cell-Free Networks} 

\author{Saeed Mashdour$^{\star}$, André R. Flores $^{\star}$, Rodrigo C. de Lamare $^{\star,\dagger}$, \vspace{-1em} 

\thanks{Saeed Mashdour, André R. Flores and  Rodrigo C. de Lamare are with Pontifical Catholic University of Rio de Janeiro, 22541-041 Rio de Janeiro, Brazil, Rodrigo C. de Lamare is with the University of York, UK. Emails are smashdour@gmail.com, \{andre.flores, delamare\}@puc-rio.br. This work was supported by CNPQ, FAPESP and FAPERJ. }}
\maketitle

\begin{abstract}
This paper presents a robust precoder design for resilient cell-free massive MIMO (CF-mMIMO) systems that minimizes the weighted sum of desired signal mean square error (MSE) and residual interference leakage power under a total transmit power constraint. The proposed robust precoder incorporates channel state information (CSI) error statistics to enhance resilience against CSI imperfections. We employ an alternating optimization algorithm initialized with a minimum MSE-type solution, which iteratively refines the precoder while maintaining low computational complexity and ensuring fast convergence. Numerical results show that the proposed method significantly outperforms conventional linear precoders, providing an effective balance between performance and computational efficiency.
\end{abstract}


\section{Introduction}

Cell‐free massive multiple-input multiple-output (CF‐mMIMO) networks have emerged as an extension of massive multiple-input multiple-output (MIMO) systems \cite{mmimo,wence} and cornerstone of next‐generation wireless systems by deploying a large number of distributed access points (APs) to jointly serve users without cell boundaries \cite{ngo2015cell, interdonato2019ubiquitous,Elhoushy2022
}.  This user‐centric architecture leverages coherent joint transmission to mitigate small‐scale fading and inter‐user interference, yielding uniformly high spectral efficiency and coverage even in cell‐edge scenarios \cite{bjornson2020scalable, ngo2017cell,Ammar2022
}.

A central challenge is how to designing resilient CF‐mMIMO networks, that employ transmit processing techniques with imperfect channel state information (CSI).  Conventional linear transmit processing schemes, such as minimum mean square error (MMSE) or zero forcing (ZF) precoders \cite{joham,gbd,wlrbd}, suffer performance degradation when CSI estimates are affected by noise, pilot contamination, or feedback delays \cite{nayebi2017precoding}. To address this, precoding schemes rely on models of imperfect CSI using the statistics of the error directly into the beamformer optimization, resulting in robust filters that remain effective despite channel estimation inaccuracies \cite{jidf,locsme,okspme,lrcc,siprec,jpba,tds,rcb,rmmseprec,siprec,rpreccf,rsprec,rscf,mbthp,rmbthp,rsthp,rrdstap,rcnn,rapa,rdiff,zcprec,rrs,mashdour2025robust}. In \cite{nayebi2017precoding}, the authors derive beamforming weights that explicitly incorporate the variance of CSI errors, yielding precoders that remain effective even when channel estimates are noisy or outdated. In \cite{palhares2021robust}, an iterative MMSE precoder was introduced that adds a loading term based on the variance of the CSI errors, making it more robust to noisy or outdated channel estimates. The work of \cite{rrs} has proposed a robust rate-splitting based precoding scheme for cell-free systems, using the MMSE criterion to mitigate multiuser interference and enhance robustness against imperfect CSI. 

In this paper, we propose a robust precoder design for resilient CF-mMIMO networks by minimizing a weighted sum of the desired-signal mean-square error and residual interference leakage power under a total transmit-power constraint. The precoder is derived in closed form, explicitly incorporating CSI error statistics to enhance robustness. Our method employs an efficient alternating-optimization algorithm, initialized with an MMSE-type solution, which refines the precoder through iterative updates. Each iteration requires only a single matrix inversion, ensuring reduced complexity, and convergence is typically achieved after a small number of iterations. Numerical results demonstrate significant performance gains over conventional MMSE and ZF precoders across varying levels of signal-to-noise ratio (SNR).

The remainder of the paper is organized as follows.  Section \ref{sec:sys_model} describes the network and channel models and the problem formulation.  Section \ref{Solut.} details the derivation of the proposed robust precoder and the alternating‐optimization algorithm.  Section \ref{Simul} presents numerical results, and Section \ref{Conclud.} concludes the paper.

We use the following notation: Bold uppercase (e.g., $\mathbf{X}$) and lowercase (e.g., $\mathbf{x}$) letters denote matrices and vectors, respectively; $\mathbf{I}_N$ is the $N \times N$ identity matrix. The complex Gaussian distribution is $\mathcal{CN}(\cdot,\cdot)$. Transpose, conjugate, and Hermitian transpose are $(\cdot)^\mathsf{T}$, $(\cdot)^*$, and $(\cdot)^\mathsf{H}$. The space of $M \times N$ complex matrices is $\mathbb{C}^{M \times N}$; $\mathrm{tr}(\cdot)$ and $\|\cdot\|_{\mathrm{F}}$ denote trace and Frobenius norm. For a vector $\mathbf{x}$, $\mathrm{diag}(\mathbf{x})$ creates a diagonal matrix, while $\mathrm{diag}(\mathbf{X})$ extracts the diagonal of $\mathbf{X}$. Expectation is $\mathbb{E}[\cdot]$, and $[\mathbf{X}]_{m,n}$ is the $(m,n)$-th entry of $\mathbf{X}$.

\section{Network and Channel Models and Problem Formulation} \label{sec:sys_model}

In this section, we detail the network architecture, the channel model adopted in this work and formulate the problem.

\subsection{Network Architecture}

We consider the downlink transmission in a CF-mMIMO network composed of $L$ geographically distributed access points (APs), each equipped with $N$ antennas. The total number of user equipment (UEs) is denoted by $K$ and is assumed to significantly exceed the number of available AP antennas, i.e., $K \gg M = LN$. To support efficient operation, only a set of $n \leq M$ UEs is scheduled for transmission per resource block, with scheduling determined by the C-ESG multiuser scheduling strategy proposed in \cite{mashdour2022enhanced}.

\subsection{Channel Representation}

The channel between the $LN$ AP antennas and the $n$ scheduled UEs is modeled as $\mathbf{G} = \hat{\mathbf{G}} + \tilde{\mathbf{G}} \in \mathbb{C}^{LN \times n}$, where $\hat{\mathbf{G}}$ denotes the estimated channel and $\tilde{\mathbf{G}}$ captures the estimation error. The elements of $\tilde{\mathbf{G}}$ follow a zero-mean complex Gaussian distribution with known variance~\cite{nayebi2017precoding}. Each element $g_{mk} = [\mathbf{G}]_{m,k}$ is expressed as:
\begin{equation} \label{eq:channel}
\begin{split}
    g_{mk} = \hat{g}_{mk} + \tilde{g}_{mk} 
    = \sqrt{1 - \alpha} \sqrt{\beta_{mk}} h_{mk} + \sqrt{\alpha} \sqrt{\beta_{mk}} \tilde{h}_{mk},
\end{split}
\end{equation}
where $\alpha \in (0,1)$ quantifies the CSI imperfection, $\beta_{mk}$ is the LSF coefficient between AP antenna $m$ and UE $k$, and $h_{mk}, \tilde{h}_{mk} \sim \mathcal{CN}(0,1)$ denotes small-scale fading and estimation error terms, respectively. These are assumed to be independent and identically distributed (i.i.d.) and statistically independent ~\cite{rscf, mishra2022rate}.

These APs collaboratively serve $n$ single antenna UEs under a user-centric CF (UCCF) approach. Each UE connects to a dynamically selected subset of APs, based on the large-scale fading (LSF) characteristics. Specifically, APs for which $\beta_{mk}$ exceeds a predefined threshold are included in the UE's serving set. This results in a sparse channel matrix $\mathbf{G}_s\in \mathbb{C}^{LN \times n}$, where only the entries corresponding to the active AP-UE pairs are nonzero.

To manage interference and improve signal quality, precoding is applied using the estimated channel matrix $\hat{\mathbf{G}}_s$ corresponding to the active AP-UE pairs. Let $\mathbf{P} \in \mathbb{C}^{LN \times n}$ denote the corresponding precoding matrix. The received signal vector at the $n$ scheduled UEs is modeled by
\begin{equation} \label{eq:rx_signal}
\begin{split}
    \mathbf{y} &= \sqrt{\rho_f} \mathbf{G}_s^T \mathbf{P} \mathbf{x} + \mathbf{w} \\
    &= \sqrt{\rho_f} \hat{\mathbf{G}}_s^T \mathbf{P} \mathbf{x} + \sqrt{\rho_f} \tilde{\mathbf{G}}_s^T \mathbf{P} \mathbf{x} + \mathbf{w},
\end{split}
\end{equation}
where $\rho_f$ is the total downlink transmit power, $\mathbf{x} \in \mathbb{C}^{n \times 1}$ is the transmit symbol vector (unit power), and $\mathbf{w} \sim \mathcal{CN}(\mathbf{0}, \sigma_w^2 \mathbf{I}_n)$ is the additive white Gaussian noise vector. The matrix $\tilde{\mathbf{G}}_s$ denotes the estimation error.

Assuming Gaussian signaling and mutual independence between $\mathbf{x}$ and $\mathbf{w}$, the achievable downlink sum-rate under imperfect CSI is upper-bounded as follows \cite{mashdour2025robust}
\begin{equation} \label{eq:sum_rate}
    \text{SR} = \log_2\left(\det\left[\mathbf{R}_{UC_{\tilde{\mathbf{G}}_s}} + \mathbf{I}_K\right]\right),
\end{equation}
where
\begin{equation}
    \mathbf{R}_{UC_{\tilde{\mathbf{G}}_s}} = \rho_f \hat{\mathbf{G}}_s^T \mathbf{P} \mathbf{P}^H \hat{\mathbf{G}}_s^* \left( \mathbf{R}_{\tilde{\mathbf{G}}_s} \right)^{-1},
\end{equation}
and
\begin{equation}
    \mathbf{R}_{\tilde{\mathbf{G}}_s} = \mathbb{E}_{\tilde{\mathbf{G}}_s} \left[ \rho_f \tilde{\mathbf{G}}_s^T \mathbf{P} \mathbf{P}^H \tilde{\mathbf{G}}_s^* \right] + \sigma_w^2 \mathbf{I}_K.
\end{equation}

\subsection{Problem Statement}
The received signal at the scheduled UEs is decomposed into
\[
\hat{\mathbf{y}}
=\sqrt{\rho_f}\,\hat{\mathbf{G}}_s^T\mathbf{P}\,\mathbf{x}+\mathbf{w},
\quad
\tilde{\mathbf{y}}
=\sqrt{\rho_f}\,\tilde{\mathbf{G}}_s^T\mathbf{P}\,\mathbf{x},
\]
where \(\hat{\mathbf{y}}\) contains the desired signal plus noise and \(\tilde{\mathbf{y}}\) captures the residual interference due to the CSI error.  We seek a precoder \(\mathbf{P}\) and scalar gain \(h\) that jointly minimize the mean square error of the desired term while penalizing the error leakage power, subject to a total transmit power constraint.
\begin{equation}\label{eq:opt_main}
\{\mathbf{P},\,h\}
=\arg\min_{\mathbf{P},\,h}\;
\underbrace{\mathbb{E}\bigl[\|\mathbf{x}-h^{-1}\hat{\mathbf{y}}\|^2\bigr]}_{\text{desired‐signal MSE}}
\;+\;
\underbrace{\mathbb{E}\bigl[\|\tilde{\mathbf{y}}\|^2\bigr]}_{\text{error‐leakage power}}
\end{equation}
\[
\text{subject to}\quad
\mathrm{tr}\bigl(\mathbf{P}^H\mathbf{P}\bigr)=P,
\]
where \(\mathbf{x}\) has unit power and \(P\) is the total downlink budget.

\section{Proposed Robust Precoding Method} \label{Solut.}

We know that $\|\mathbf{x} - h^{-1} \hat{\mathbf{y}} \|^2$ and $\| \tilde{\mathbf{y}} \|^2$ are scalars. Therefore, we can express them using the trace operator without changing their values. Hence, we write
\begin{equation}
\begin{split}
    &\|\mathbf{x} - h^{-1} \hat{\mathbf{y}} \|^2 = 
    \mathrm{tr} \left[ (\mathbf{x} - h^{-1} \hat{\mathbf{y}})^H (\mathbf{x} - h^{-1} \hat{\mathbf{y}}) \right]
    \end{split},
\end{equation}
where $\mathbf{x} - h^{-1} \hat{\mathbf{y}}=\mathbf{x} - h^{-1} \sqrt{\rho_f} \hat{\mathbf{G}}_s^T \mathbf{P} \mathbf{x} - h^{-1} \mathbf{w}$. 
The expectation of this expression is given by
\begin{equation}
\begin{split}
    &\mathbb{E} \left[ \| \mathbf{x} - h^{-1} \hat{\mathbf{y}} \|^2 \right] = \\
    &\mathbb{E} \left[ \text{tr} \left[ \left( \mathbf{x} - h^{-1} \hat{\mathbf{y}} \right)^H \left( \mathbf{x} - h^{-1} \hat{\mathbf{y}} \right) \right] \right]=\\
    ~& \text{tr} \left( \mathbb{E} \left[ \left( \mathbf{x} - h^{-1} \hat{\mathbf{y}} \right)^H \left( \mathbf{x} - h^{-1} \hat{\mathbf{y}} \right) \right] \right) =\\
    &\text{tr} \left( \mathbb{E} \left[ \mathbf{x}^H \mathbf{x} \right] \right) 
- \text{tr} \left(\mathbb{E} \left[ h^{-1} \mathbf{x}^H \sqrt{\rho_f} \hat{\mathbf{G}}_s^T \mathbf{P} \mathbf{x} \right] \right) 
- \\
&  \text{tr} \left(\mathbb{E} \left[ h^{-1} \mathbf{x}^H \mathbf{w} \right]\right) + \text{tr} \left(\mathbb{E} \left[ \sqrt{\rho_f} h^{-1} \mathbf{x}^H \mathbf{P}^H \hat{\mathbf{G}}_s^* \right] \right)
+ \\
& \text{tr} \left(\mathbb{E} \left[ \rho_f h^{-2} \mathbf{x}^H \mathbf{P}^H \hat{\mathbf{G}}_s^* \hat{\mathbf{G}}_s^T \mathbf{P} \mathbf{x} \right]\right)+   \\
& \text{tr} \left(\mathbb{E}\left[ h^{-2} \sqrt{\rho_f} \mathbf{x}^H \mathbf{P}^H \hat{\mathbf{G}}_s^* \mathbf{w} \right]\right)
- \\
&\text{tr} \left(\mathbb{E} \left[ h^{-1} \mathbf{w}^H \mathbf{x} \right]\right) + \text{tr} \left(\mathbb{E}\left[ h^{-2} \sqrt{\rho_f} \mathbf{w}^H  \hat{\mathbf{G}}_s^T \mathbf{P}\mathbf{x} \right]\right)
+ \\
& \text{tr} \left(\mathbb{E} \left[ h^{-2} \mathbf{w}^H \mathbf{w} \right]\right)  = \text{tr}(\mathbf{C}_x)-\text{tr}\left( h^{-1} \sqrt{\rho_f} \hat{\mathbf{G}}_s^T\mathbf{P}  \right)
-  \\
& \text{tr}\left( h^{-1} \sqrt{\rho_f} \mathbf{P}^H \hat{\mathbf{G}}_s^* \right)
+ \text{tr}\left( h^{-2} \rho_f \mathbf{P}^H \hat{\mathbf{G}}_s^* \hat{\mathbf{G}}_s^T \mathbf{P} \right)
+ \text{tr}(h^{-2} \mathbf{C}_w).
\end{split} 
\end{equation}

Since $\mathbf{C}_x = \mathbf{I}_n$ and $\mathbf{C}_w = \sigma_w^2 \mathbf{I}_n$, and $\mathbf{w}$, $\mathbf{x}$ are zero mean and uncorrelated, we will have:

\begin{equation}
    \begin{split}  
\mathbb{E} \left[ \| \mathbf{x} - h^{-1} \hat{\mathbf{y}} \|^2 \right]
&= n - \text{tr} \left( h^{-1} \sqrt{\rho_f} \hat{\mathbf{G}}_s^T \mathbf{P} \right)
   \\
& - \text{tr} \left( h^{-1} \sqrt{\rho_f} \mathbf{P}^H \hat{\mathbf{G}}_s^* \right) \\
& + \text{tr} \left( h^{-2} \rho_f \mathbf{P}^H \hat{\mathbf{G}}_s^* \hat{\mathbf{G}}_s^T \mathbf{P} \right)
+ h^{-2} \sigma_w^2 n.
\end{split}
\end{equation}

Now, for $\mathbb{E} \left[ \| \tilde{\mathbf{y}} \|^2 \right]$ we have the following:

\begin{equation}
  \begin{split} 
\mathbb{E} \left[ \| \tilde{\mathbf{y}} \|^2 \right] 
&= \mathbb{E} \left[ \tilde{\mathbf{y}}^H \tilde{\mathbf{y}} \right] 
= \text{tr} \left( \mathbb{E} \left[ \tilde{\mathbf{y}}^H \tilde{\mathbf{y}} \right] \right) \\
&= \text{tr} \left( \mathbb{E} \left[  \left( \sqrt{\rho_f} \tilde{\mathbf{G}}_s^T \mathbf{P} \mathbf{x} \right)^H \cdot \sqrt{\rho_f} \tilde{\mathbf{G}}_s^T \mathbf{P} \mathbf{x} \right] \right) \\
&= \text{tr} \left( \mathbb{E} \left[   \sqrt{\rho_f} \mathbf{x} ^H \mathbf{P}^H \tilde{\mathbf{G}}_s^*  \sqrt{\rho_f} \tilde{\mathbf{G}}_s^T \mathbf{P} \mathbf{x} \right] \right) \\
&= \text{tr} \left( \rho_f  \mathbf{P}^H \mathbb{E} \left[ \tilde{\mathbf{G}}_s^* \tilde{\mathbf{G}}_s^T \right] \mathbf{P}  \right)
\end{split}  
\end{equation}

Thus, the objective function in the optimization problem \eqref{eq:opt_main} is:
\begin{equation}
    \begin{split}
        J &= n + h^{-2} \sigma_w^2 n 
- \text{tr} \left( h^{-1} \sqrt{\rho_f} \hat{\mathbf{G}}_s^T \mathbf{P} \right) \\ 
& - \text{tr} \left( h^{-1} \sqrt{\rho_f} \mathbf{P}^H \hat{\mathbf{G}}_s^* \right) + \text{tr} \left( h^{-2} \rho_f \mathbf{P}^H \hat{\mathbf{G}}_s^* \hat{\mathbf{G}}_s^T \mathbf{P} \right) \\
& + \text{tr} \left( \rho_f  \mathbf{P}^H \mathbb{E} \left[ \tilde{\mathbf{G}}_s^* \tilde{\mathbf{G}}_s^T \right] \mathbf{P} \right).
    \end{split}
\end{equation}

Then, the Lagrangian of the problem \eqref{eq:opt_main} is given by
\begin{equation}
    \begin{split}
         \mathcal{L}\left ( \mathbf{P}, h, \lambda  \right )& = n + h^{-2} \sigma_w^2 n 
- \text{tr} \left( h^{-1} \sqrt{\rho_f} \hat{\mathbf{G}}_s^T \mathbf{P} \right) 
 \\ &  - \text{tr} \left( h^{-1} \sqrt{\rho_f} \mathbf{P}^H \hat{\mathbf{G}}_s^* \right) + \text{tr} \left( h^{-2} \rho_f \mathbf{P}^H \hat{\mathbf{G}}_s^* \hat{\mathbf{G}}_s^T \mathbf{P} \right) \\
 & + \text{tr} \left( \rho_f  \mathbf{P}^H \mathbb{E} \left[ \tilde{\mathbf{G}}_s^* \tilde{\mathbf{G}}_s^T \right] \mathbf{P} \right)+\lambda\left (  \mathrm{tr}(\mathbf{P}^H \mathbf{P} )-P \right ).
    \end{split}
\end{equation}
We now take the derivative of the Lagrangian with respect to the precoding matrix $\mathbf{P}$. We use the property $\frac{\partial \, \mathrm{tr}(\mathbf{A}\mathbf{Z})}{\partial \mathbf{Z}} = \mathbf{A}^T$ \cite{hjorungnes2007complex} and obtain
\begin{equation}
    \frac{\partial \, \text{tr} \left( h^{-1} \sqrt{\rho_f} \hat{\mathbf{G}}_s^T \mathbf{P} \right)}{\partial \mathbf{P}} =  h^{-1} \sqrt{\rho_f} \hat{\mathbf{G}}_s.
\end{equation}
Using the property $\frac{\partial \, \mathrm{tr}(\mathbf{Z}^H\mathbf{A})}{\partial \mathbf{Z}} = \mathbf{0}$, we obtain

\begin{equation}
    \frac{\partial \, \text{tr} \left( h^{-1} \sqrt{\rho_f} \mathbf{P}^H \hat{\mathbf{G}}_s^* \right)}{\partial \mathbf{P}} =  \mathbf{0}.
\end{equation}
Using the property $\frac{\partial}{\partial \mathbf{Z}}\operatorname{tr}\left\{ \mathbf{Z} \mathbf{A}_0 \mathbf{Z}^H \mathbf{A}_1 \right\}
=\mathbf{A}_1^T \mathbf{Z}^* \mathbf{A}_0^T$, the following are derived
\begin{equation}
    \frac{\partial \, \text{tr} \left( h^{-2} \rho_f \mathbf{P}^H \hat{\mathbf{G}}_s^* \hat{\mathbf{G}}_s^T \mathbf{P} \right)}{\partial \mathbf{P}} =  h^{-2} {\rho_f} \hat{\mathbf{G}}_s \hat{\mathbf{G}}_s^H \mathbf{P}^*,
\end{equation}
\begin{equation}
    \frac{\partial \, \text{tr} \left( \rho_f  \mathbf{P}^H \mathbb{E} \left[ \tilde{\mathbf{G}}_s^* \tilde{\mathbf{G}}_s^T \right] \mathbf{P} \right)}{\partial \mathbf{P}} =  \rho_f \mathbb{E} \left[ \tilde{\mathbf{G}}_s \tilde{\mathbf{G}}_s^H \right] \mathbf{P}^*,
\end{equation}
\begin{equation}
    \frac{\partial \, \mathrm{tr}(\mathbf{P}^H \mathbf{P} )}{\partial \mathbf{P}} =  \mathbf{P}^*.
\end{equation}
Thus, the derivative of the Lagrangian with respect to $\mathbf{P}$ is obtained as follows:
\begin{equation}
\begin{split}
    \frac{\partial \, \mathcal{L}\left ( \mathbf{P}, h, \lambda  \right )}{\partial \mathbf{P}} &=  -h^{-1} \sqrt{\rho_f} \hat{\mathbf{G}}_s+h^{-2} {\rho_f} \hat{\mathbf{G}}_s \hat{\mathbf{G}}_s^H \mathbf{P}^*\\
    &+\rho_f \mathbb{E} \left[ \tilde{\mathbf{G}}_s \tilde{\mathbf{G}}_s^H \right] \mathbf{P}^*+\lambda \mathbf{P}^*.
    \end{split}
\end{equation}
Equating the derivative of the Lagrangian to zero, we have
\begin{equation}
    -h^{-1}\sqrt{\rho_f}\,\hat{\mathbf{G}}_s + \Bigl( h^{-2}\rho_f\,\hat{\mathbf{G}}_s\hat{\mathbf{G}}_s^H + \rho_f\,\mathbb{E}\bigl[\tilde{\mathbf{G}}_s\tilde{\mathbf{G}}_s^H\bigr] + \lambda \mathbf{I} \Bigr) \mathbf{P}^* = \mathbf{0}.
\end{equation}
We then rearrange the equation so that all terms involving $\mathbf{P}^*$ are on one side as follows
\begin{equation} \label{eq.P}
    \Bigl( h^{-2}\rho_f\,\hat{\mathbf{G}}_s\hat{\mathbf{G}}_s^H + \rho_f\,\mathbb{E}\bigl[\tilde{\mathbf{G}}_s\tilde{\mathbf{G}}_s^H\bigr] + \lambda \mathbf{I} \Bigr) \mathbf{P}^* = h^{-1}\sqrt{\rho_f}\,\hat{\mathbf{G}}_s.
\end{equation}
By defining the matrix $\mathbf{A} = \rho_f\,\hat{\mathbf{G}}_s\hat{\mathbf{G}}_s^H + h^{2} \rho_f\,\mathbb{E}\bigl[\tilde{\mathbf{G}}_s\tilde{\mathbf{G}}_s^H\bigr] + h^{2} \lambda \mathbf{I}$, and assuming it invertible, we can rewrite $\mathbf{P}^* = h\sqrt{\rho_f}\mathbf{A}^{-1}\hat{\mathbf{G}}_s$. Thus, the optimal precoder matrix is obtained as
\begin{equation}  \label{P_Opt} 
\begin{split}
    \mathbf{P}_{opt} =& h \rho_f (\mathbf{A}^*)^{-1} \hat{\mathbf{G}}_{s}^*=\\
    &
    h \rho_f\left(  \rho_f \hat{\mathbf{G}}_s^{*} \hat{\mathbf{G}}_s^{T} + h^{2} \rho_f \mathbb{E} \left[ \tilde{\mathbf{G}}_s^{*} \tilde{\mathbf{G}}_s^{T} \right] + h^{2}\lambda \mathbf{I} \right)^{-1} \hat{\mathbf{G}}_{s}^*.
\end{split}
\end{equation}
Taking the derivative of the Lagrangian with respect to $h$, we obtain
\begin{equation} \label{deriv.h}
    \begin{split}
        \frac{\partial \mathcal{L}(\mathbf{P}, h, \lambda)}{\partial h} 
=& -2 h^{-3} \sigma_w^2 n 
+ h^{-2} \sqrt{\rho_f} \, \text{tr} \left( \hat{\mathbf{G}}_s^T \mathbf{P} \right)+  \\
&
h^{-2} \sqrt{\rho_f} \, \text{tr} \left( \mathbf{P}^H \hat{\mathbf{G}}_s^* \right) 
- \\
&2 h^{-3} \rho_f \, \text{tr} \left( \mathbf{P}^H \hat{\mathbf{G}}_s^* \hat{\mathbf{G}}_s^T \mathbf{P} \right).
    \end{split}
\end{equation}
Equating \eqref{deriv.h} to zero, we obtain the following
\begin{equation} \label{deriv1.h}
    \begin{split}
        h \sqrt{\rho_f} \, \text{Re}\{\text{tr}(\hat{\mathbf{G}}_s^T \mathbf{P})\} = \sigma_w^2 n - \rho_f \, \text{tr}(\mathbf{P}^H \hat{\mathbf{G}}_s^* \hat{\mathbf{G}}_s^T \mathbf{P}).
    \end{split}
\end{equation}
We can also rewrite \eqref{eq.P} as follows
\begin{equation} \label{new.eq.p}
     h\sqrt{\rho_f}\,\hat{\mathbf{G}}_s^T= \mathbf{P}^H\Bigl( \rho_f\,\hat{\mathbf{G}}_s^*\hat{\mathbf{G}}_s^T + h^2\rho_f\,\mathbb{E}\bigl[\tilde{\mathbf{G}}_s^*\tilde{\mathbf{G}}_s^T\bigr] + \lambda h^2 \mathbf{I} \Bigr).
\end{equation}
From \eqref{new.eq.p} and $\mathrm{tr}(\mathbf{P}^H \mathbf{P} ) = P$ we obtain
\begin{equation} \label{new1.eq.p}
    \begin{split}
        h \sqrt{\rho_f} \cdot \mathrm{tr}(\hat{\mathbf{G}}_s^T \mathbf{P}) =& \rho_f \, \mathrm{tr}(\mathbf{P}^H \hat{\mathbf{G}}_s^*\hat{\mathbf{G}}_s^T \mathbf{P}) +\\
        &h^2 \rho_f \, \mathrm{tr}(\mathbf{P}^H \mathbb{E}[\tilde{\mathbf{G}}_s^*\tilde{\mathbf{G}}_s^T] \mathbf{P}) + \lambda h^2 P.
    \end{split}
\end{equation}
Using \eqref{deriv1.h} and \eqref{new1.eq.p}, we can obtain $\lambda$. Taking the real part of \eqref{new1.eq.p}, since the trace of a Hermitian matrix is always real, we have

\begin{equation} \label{eq2_real}
\begin{split}
h \sqrt{\rho_f} \, \text{Re}\{\text{tr}(\hat{\mathbf{G}}_s^T \mathbf{P})\} = &\rho_f \, \text{tr}(\mathbf{P}^H \hat{\mathbf{G}}_s^*\hat{\mathbf{G}}_s^T \mathbf{P}) + \\
&h^2 \rho_f \, \text{tr}(\mathbf{P}^H \mathbb{E}[\tilde{\mathbf{G}}_s^*\tilde{\mathbf{G}}_s^T] \mathbf{P}) + \lambda h^2 P_t .
\end{split}
\end{equation}
Substituting \eqref{deriv1.h} into \eqref{eq2_real}, we obtain
\begin{equation}
    \begin{split}   
\sigma_w^2 n - \rho_f \, \text{tr}(\mathbf{P}^H \hat{\mathbf{G}}_s^* \hat{\mathbf{G}}_s^T \mathbf{P}) &= \rho_f \, \text{tr}(\mathbf{P}^H \hat{\mathbf{G}}_s^*\hat{\mathbf{G}}_s^T \mathbf{P}) \\
& + h^2 \rho_f \, \text{tr}(\mathbf{P}^H \mathbb{E}[\tilde{\mathbf{G}}_s^*\tilde{\mathbf{G}}_s^T] \mathbf{P})  \\
& + \lambda h^2 P.
\end{split}
\end{equation}
Solving for $\lambda$, we obtain

\begin{equation}
  \begin{split}
\lambda h^2 P_t = &\sigma_w^2 n - 2\rho_f \, \text{tr}(\mathbf{P}^H \hat{\mathbf{G}}_s^* \hat{\mathbf{G}}_s^T \mathbf{P}) -\\
&h^2 \rho_f \, \text{tr}(\mathbf{P}^H \mathbb{E}[\tilde{\mathbf{G}}_s^*\tilde{\mathbf{G}}_s^T] \mathbf{P}).
\end{split}  
\end{equation}
Thus
\begin{equation}
    \begin{split}
        \lambda &= \frac{\sigma_w^2 n}{h^2 P} - \frac{2\rho_f \, \text{tr}(\mathbf{P}^H \hat{\mathbf{G}}_s^* \hat{\mathbf{G}}_s^T \mathbf{P})}{h^2 P} - \frac{\rho_f \, \text{tr}(\mathbf{P}^H \mathbb{E}[\tilde{\mathbf{G}}_s^*\tilde{\mathbf{G}}_s^T] \mathbf{P})}{P}
    \end{split}
\end{equation}
In order to obtain $h$, we use $\mathrm{tr}(\mathbf{P}^H \mathbf{P} ) = P$ and $\mathbf{P}_{opt}$ as derived in \eqref{P_Opt}. Defining $\mathbf{B}=(\mathbf{A}^*)^{-1} \hat{\mathbf{G}}_{s}^*$, we rewrite $\mathbf{P} = h \sqrt{\rho_f} \mathbf{B}$ into the power constraint as
\begin{equation}
    \mathrm{tr}\left(\mathbf{P}^H\mathbf{P}\right) = h^2 \rho_f \, \mathrm{tr}\left(\mathbf{B}^H\mathbf{B}\right) = P.
\end{equation}
Thus, we obtain $h$ as
\begin{equation}
h = \frac{1}{\sqrt{\rho_f}} \sqrt{\frac{P}{\mathrm{tr}\left(\mathbf{B}^H\mathbf{B}\right)}}.
\end{equation}

Since the interdependence between the precoder $\mathbf{P}$ and the regularization parameter $\lambda$ makes simultaneous optimization challenging, we adopt an optimization framework in which one variable is held fixed while the other is updated, and the process is repeated until convergence. We begin with a standard MMSE precoder as an initial guess, which aligns beamforming with the dominant channel subspace. This initialization already incorporates noise and interference statistics, yielding a starting point that accelerates convergence and follows well-established robust design practices. This procedure is summarized in Algorithm \ref{alg:robust_precoder}.


\begin{algorithm}[t]
\caption{Proposed Robust Precoder}\label{alg:robust_precoder}
\KwIn{$\hat{\mathbf{G}}_s,\;\rho_f,\;P,\;\sigma_w^2,\;\Psi \;=\;\mathbb E[\tilde G_s^*\,\tilde G_s^T],\;\varepsilon,\;I_{\max}$}
\KwOut{Precoder $\mathbf{P}$, scaling $h$, regularizer $\lambda$}

\SetKwBlock{Init}{\textbf{Initialization:}}{}
\Init{
  \[
    \mathbf{B}^{(0)}
    \leftarrow
    \Bigl(\rho_f\,\hat{\mathbf{G}}_s\,\hat{\mathbf{G}}_s^T
      + \tfrac{\sigma_w^2\,n}{P}\,\mathbf{I}
    \Bigr)^{-1}\,\hat{\mathbf{G}}_s^*,
  \]
  \[
    h^{(0)}
    \leftarrow
    \frac{1}{\sqrt{\rho_f}}
    \sqrt{\frac{P}{\mathrm{tr}\bigl(\mathbf{B}^{(0)H}\mathbf{B}^{(0)}\bigr)}} ,
    \quad
    \mathbf{P}^{(0)}
    \leftarrow
    h^{(0)}\,\sqrt{\rho_f}\,\mathbf{B}^{(0)},
  \]
  \[
    \lambda^{(0)}
    \leftarrow
    \frac{\sigma_w^2\,n}{(h^{(0)})^2\,P}
    \;-\;\frac{2\,\rho_f}{(h^{(0)})^2\,P}\,
      \mathrm{tr}\bigl(\mathbf{P}^{(0)H}\,\hat{\mathbf{G}}_s\,
      \hat{\mathbf{G}}_s^T\,\mathbf{P}^{(0)}\bigr)
  \]
  \[
    \hspace{4em}
    -\;\frac{\rho_f}{P}\,
      \mathrm{tr}\bigl(\mathbf{P}^{(0)H}\,\Psi\,\mathbf{P}^{(0)}\bigr).
  \]
}

\For{$i\leftarrow 1$ \KwTo $I_{\max}$}{
  

  \[
  \mathbf{M}^{(i)}
  = \rho_f\,\hat{\mathbf{G}}_s\,\hat{\mathbf{G}}_s^T
    + (h^{(i-1)})^2\,\rho_f\,\Psi
    + (h^{(i-1)})^2\,\lambda^{(i-1)}\,\mathbf{I}.
\]

\[
  B^{(i)}
  \leftarrow
  \bigl(\mathbf{M}^{(i)}\bigr)^{-1}\,\hat{G}_s^*.
\]
  
  \[
    h^{(i)}
    \leftarrow
    \frac{1}{\sqrt{\rho_f}}
    \sqrt{\frac{P}{\mathrm{tr}\bigl(\mathbf{B}^{(i)H}\mathbf{B}^{(i)}\bigr)}} ,
    \quad
    \mathbf{P}^{(i)}
    \leftarrow
    h^{(i)}\,\sqrt{\rho_f}\,\mathbf{B}^{(i)}.
  \]
  
  \[
    \lambda^{(i)}
    \leftarrow
    \frac{\sigma_w^2\,n}{(h^{(i)})^2\,P}
    \;-\;\frac{2\,\rho_f}{(h^{(i)})^2\,P}\,
      \mathrm{tr}\bigl(\mathbf{P}^{(i)H}\,\hat{\mathbf{G}}_s\,
      \hat{\mathbf{G}}_s^T\,\mathbf{P}^{(i)}\bigr)
  \]
  \[
    \hspace{4em}
    -\;\frac{\rho_f}{P}\,
      \mathrm{tr}\bigl(\mathbf{P}^{(i)H}\,\Psi\,\mathbf{P}^{(i)}\bigr).
  \]
}

\Return{$\mathbf{P}\!\leftarrow P^{(end)},\;h\!\leftarrow h^{(end)},\;\lambda\!\leftarrow \lambda^{(end)}$}
\end{algorithm}

\section{Computational Complexity Analysis}
The computational complexity of the proposed robust precoder is mainly driven by matrix operations, iterative updates, and matrix inversions. The construction of the matrix $\mathbf{M}^{(i-1)}$ involves matrix multiplications, resulting in a complexity of $\mathcal{O}(n(LN)^2)$, where $M = LN$ is the total number of transmit antennas and $n$ is the number of scheduled users. The matrix inversion step for the update of $\mathbf{M}^{(i-1)}$ has a complexity of $\mathcal{O}(n^3)$, as it requires the inversion of an $n \times n$ matrix. The matrix-vector multiplications to compute $\mathbf{B}^{(i)}$ and update the precoder matrix $\mathbf{P}^{(i)}$ at each iteration have a complexity of $\mathcal{O}(n(LN)^2)$. Additionally, the calculation of the regularization parameter $\lambda^{(i)}$ involves trace operations, which contribute $\mathcal{O}(n^2)$ to the complexity. Consequently, the total complexity per iteration is $\mathcal{O}(n^3 + n(LN)^2)$. Since the algorithm runs for $I_{\max}$ iterations, the overall computational complexity of the proposed robust precoder is $\mathcal{O}(I_{\max} \cdot (n^3 + n(LN)^2))$.  In comparison, conventional MMSE and ZF precoders require $\mathcal{O}(n^3 + n(LN)^2)$ operations for their single matrix inversion and multiplication steps. Thus, while the proposed robust precoder has higher complexity due to its iterative nature, it provides robustness to channel estimation errors, making it more suitable for scenarios where this robustness is essential. In contrast, the ZF and MMSE precoders, though computationally less demanding, do not offer the same level of robustness to imperfect CSI.

\section{Numerical Results} \label{Simul}
In this section, we evaluate the performance of the proposed robust precoder in a downlink user-centric CF-mMIMO network. The network consists of $L = 16$ APs, each equipped with $N = 4$ antennas, resulting in a total of $M = LN = 64$ transmit antennas. The system serves $K = 128$ single-antenna UEs that are randomly distributed across a square area of 400 meters by 400 meters. Since the number of users is much greater than the number of antennas, we employ the C-ESG user scheduling algorithm \cite{mashdour2022enhanced} to select $n = 16$ users for each transmission block. The AP selection is based on LSF. Equal power loading is used for all UEs in the system, and Gaussian signaling is assumed.

\begin{figure}[t] \centering \includegraphics[width=1\linewidth]{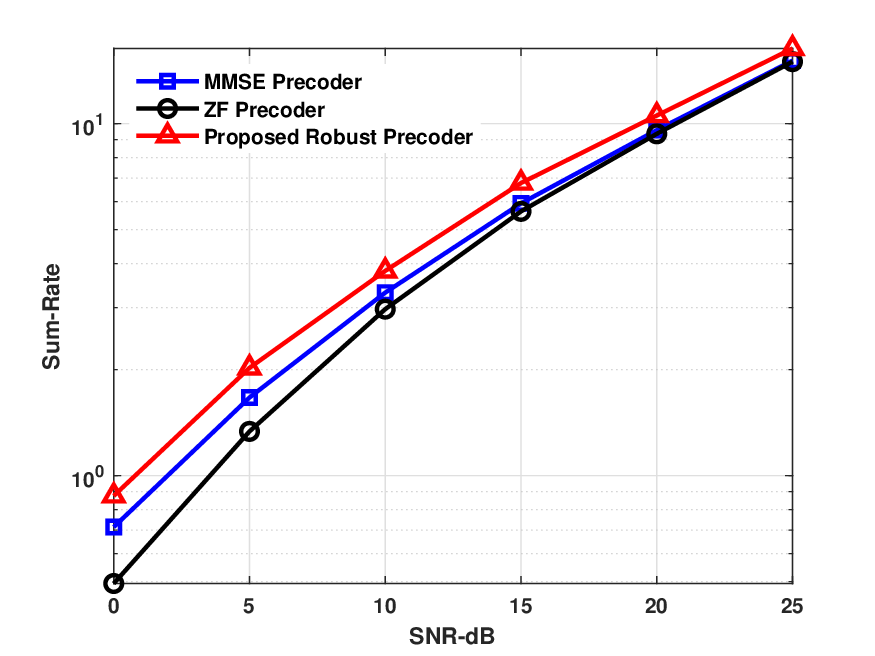} \vspace{-0.75em} \caption{\small{Sum-rate comparison of the proposed robust precoder, ZF, and MMSE with equal power loading and $\alpha=0.15$, $L = 16$, $N = 4$, $K = 128$, $n = 16$, and C-ESG scheduling.}} \vspace{-0.5em} \label{fig:fig1} \end{figure}

Fig.~\ref{fig:fig1} compares the sum-rate performance of the proposed robust precoder with that of the ZF and MMSE precoders. The results show that while MMSE outperforms ZF, the proposed method exhibits superior performance over both ZF and MMSE as SNR increases. This demonstrates that the robust precoder is effective and resilient to channel estimation errors, outperforming both conventional precoding schemes under the considered channel conditions.

\begin{figure}[t] \centering \includegraphics[width=1\linewidth]{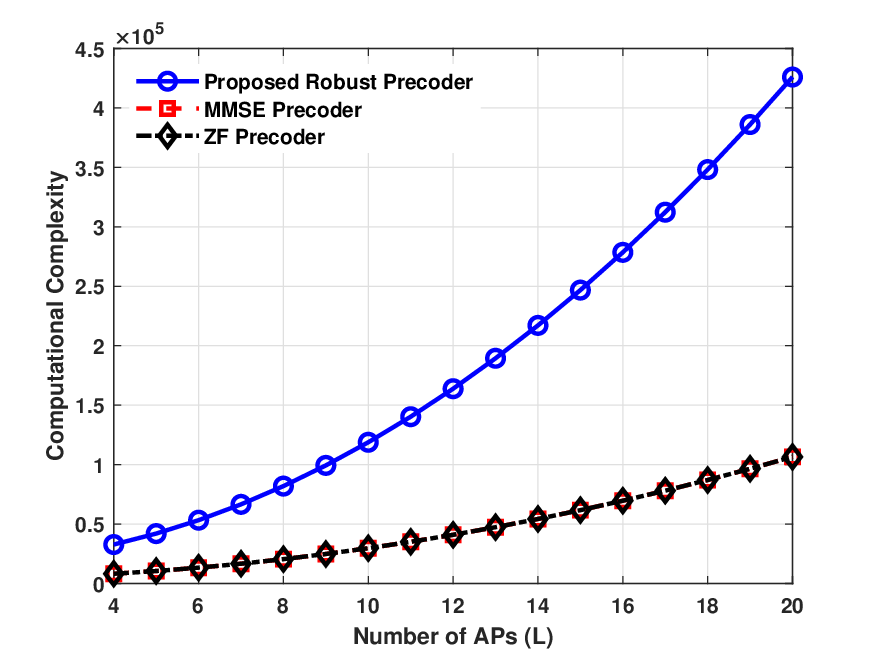} \vspace{-0.75em} \caption{\small{Computational complexity comparison of the proposed robust precoder, ZF, and MMSE in terms of floating point operations, when $n = 16$ UEs are scheduled.}} \vspace{-0.5em} \label{fig:fig2} \end{figure}

Fig.~\ref{fig:fig2} presents the comparison of the computational complexities of the proposed robust precoder with 4 iterations, ZF, and MMSE methods. The proposed method has higher computational complexity due to its iterative nature, but this additional complexity is outweighed by the substantial improvements in sum-rate performance across different values of SNR. This shows that the proposed robust precoder provides an effective balance between performance and computational efficiency, making it a valuable solution for large-scale CF-mMIMO systems.

\section{Conclusions} \label{Conclud.}
In this paper, we introduced a robust precoder design for resilient CF-mMIMO networks that minimizes the desired signal MSE and the residual interference leakage power while adhering to a total transmit power constraint. Our method, which incorporates CSI error statistics, significantly enhances the system's robustness against channel estimation errors. The proposed alternating optimization algorithm has low complexity, with each iteration requiring only a single matrix inversion and convergence in a small number of iterations. Numerical results have shown that the proposed precoder outperforms both ZF and MMSE precoders across different SNR values. Although the proposed method introduces higher computational complexity due to its iterative nature, the performance improvements in sum-rate outweigh this cost, providing an effective balance between performance and computational efficiency. This makes our method a valuable approach for resilient CF-mMIMO networks.

\bibliographystyle{IEEEbib}
\bibliography{refs}

\begin{thebibliography}{10}

\bibitem{mmimo}
R.~C. de~Lamare,
\newblock ``Massive mimo systems: Signal processing challenges and future trends,''
\newblock {\em URSI Radio Science Bulletin}, vol. 2013, no. 347, pp. 8--20, 2013.

\bibitem{wence}
W.~Zhang, H.~Ren, C.~Pan, M.~Chen, R.~C. de~Lamare, B.~Du, and J.~Dai,
\newblock ``Large-scale antenna systems with ul/dl hardware mismatch: Achievable rates analysis and calibration,''
\newblock {\em IEEE Transactions on Communications}, vol. 63, no. 4, pp. 1216--1229, 2015.

\bibitem{ngo2015cell}
H.~Q. Ngo, A.~Ashikhmin, H.~Yang, E.~G. Larsson, and T.~L. Marzetta,
\newblock ``Cell-free massive {MIMO}: Uniformly great service for everyone,''
\newblock IEEE 16th International Workshop on Signal Processing Advances in Wireless Communications (SPAWC), 2015.

\bibitem{interdonato2019ubiquitous}
G.~Interdonato, E.~Björnson, H.~Q. Ngo, P.~Frenger, and E.~G. Larsson,
\newblock ``Ubiquitous cell-free massive {MIMO} communications,''
\newblock {\em EURASIP Journal on Wireless Communications and Networking}, vol. 2019, no. 1, pp. 1--13, 2019.

\bibitem{Elhoushy2022}
Salah Elhoushy, Mohamed Ibrahim, and Walaa Hamouda,
\newblock ``Cell-free massive {MIMO}: A survey,''
\newblock {\em IEEE Communications Surveys \& Tutorials}, vol. 24, no. 1, pp. 492--523, 2022.

\bibitem{bjornson2020scalable}
E.~Björnson and L.~Sanguinetti,
\newblock ``Scalable cell-free massive {MIMO} systems,''
\newblock {\em IEEE Transactions on Communications}, vol. 68, no. 7, pp. 4247--4261, 2020.

\bibitem{ngo2017cell}
H.~Q. Ngo, A.~Ashikhmin, H.~Yang, E.~G. Larsson, and T.~L. Marzetta,
\newblock ``Cell-free massive {MIMO} versus small cells,''
\newblock {\em IEEE Transactions on Wireless Communications}, vol. 16, no. 3, pp. 1834--1850, 2017.

\bibitem{Ammar2022}
Hussein~A. Ammar, Raviraj Adve, Shahram Shahbazpanahi, Gary Boudreau, and Kothapalli~Venkata Srinivas,
\newblock ``User-centric cell-free massive {MIMO} networks: A survey of opportunities, challenges and solutions,''
\newblock {\em IEEE Communications Surveys \& Tutorials}, vol. 24, no. 1, pp. 611--652, 2022.

\bibitem{joham}
M.~Joham, W.~Utschick, and J.A. Nossek,
\newblock ``Linear transmit processing in mimo communications systems,''
\newblock {\em IEEE Transactions on Signal Processing}, vol. 53, no. 8, pp. 2700--2712, 2005.

\bibitem{gbd}
K.~Zu, R.~C. de~Lamare, and M.~Haardt,
\newblock ``Generalized design of low-complexity block diagonalization type precoding algorithms for multiuser mimo systems,''
\newblock {\em IEEE Transactions on Communications}, vol. 61, no. 10, pp. 4232--4242, 2013.

\bibitem{wlrbd}
W.~Zhang, R.~C. de~Lamare, C.~Pan, M.~Chen, J.~Dai, B.~Wu, and X.~Bao,
\newblock ``Widely linear precoding for large-scale mimo with iqi: Algorithms and performance analysis,''
\newblock {\em IEEE Transactions on Wireless Communications}, vol. 16, no. 5, pp. 3298--3312, 2017.

\bibitem{nayebi2017precoding}
E.~Nayebi, A.~Ashikhmin, T.~L. Marzetta, H.~Yang, and B.~D. Rao,
\newblock ``Precoding and power optimization in cell-free massive {MIMO} systems,''
\newblock {\em IEEE Transactions on Wireless Communications}, vol. 16, no. 7, pp. 4445--4459, 2017.

\bibitem{jidf}
R.~C. de~Lamare and R.~Sampaio-Neto,
\newblock ``Adaptive reduced-rank processing based on joint and iterative interpolation, decimation, and filtering,''
\newblock {\em IEEE Transactions on Signal Processing}, vol. 57, no. 7, pp. 2503--2514, 2009.

\bibitem{locsme}
H.~Ruan and R.~C. de~Lamare,
\newblock ``Robust adaptive beamforming using a low-complexity shrinkage-based mismatch estimation algorithm,''
\newblock {\em IEEE Signal Processing Letters}, vol. 21, no. 1, pp. 60--64, 2014.

\bibitem{okspme}
H.~Ruan and R.~C. de~Lamare,
\newblock ``Robust adaptive beamforming based on low-rank and cross-correlation techniques,''
\newblock {\em IEEE Transactions on Signal Processing}, vol. 64, no. 15, pp. 3919--3932, 2016.

\bibitem{lrcc}
H.~Ruan and R.~C. de~Lamare,
\newblock ``Distributed robust beamforming based on low-rank and cross-correlation techniques: Design and analysis,''
\newblock {\em IEEE Transactions on Signal Processing}, vol. 67, no. 24, pp. 6411--6423, 2019.

\bibitem{siprec}
Y.~Cai, R.~C.~de Lamare, and R.~Fa,
\newblock ``Switched interleaving techniques with limited feedback for interference mitigation in ds-cdma systems,''
\newblock {\em IEEE Transactions on Communications}, vol. 59, no. 7, pp. 1946--1956, 2011.

\bibitem{jpba}
Y.~Jiang, Y.~Zou, H.~Guo, T.~A. Tsiftsis, M.~R. Bhatnagar, R.~C. de~Lamare, and Y.-D. Yao,
\newblock ``Joint power and bandwidth allocation for energy-efficient heterogeneous cellular networks,''
\newblock {\em IEEE Transactions on Communications}, vol. 67, no. 9, pp. 6168--6178, 2019.

\bibitem{tds}
P.~Clarke and R.~C. de~Lamare,
\newblock ``Transmit diversity and relay selection algorithms for multirelay cooperative mimo systems,''
\newblock {\em IEEE Transactions on Vehicular Technology}, vol. 61, no. 3, pp. 1084--1098, 2012.

\bibitem{rcb}
S.~D. Somasundaram, N.~H. Parsons, P.~Li, and R.~C. de~Lamare,
\newblock ``Reduced-dimension robust capon beamforming using krylov-subspace techniques,''
\newblock {\em IEEE Transactions on Aerospace and Electronic Systems}, vol. 51, no. 1, pp. 270--289, 2015.

\bibitem{rmmseprec}
Y.~Cai, R.~C. de~Lamare, L.-L. Yang, and M.~Zhao,
\newblock ``Robust mmse precoding based on switched relaying and side information for multiuser mimo relay systems,''
\newblock {\em IEEE Transactions on Vehicular Technology}, vol. 64, no. 12, pp. 5677--5687, 2015.

\bibitem{rpreccf}
V.~M.~T. Palhares, A.~R. Flores, and R.~C. de~Lamare,
\newblock ``Robust mmse precoding and power allocation for cell-free massive mimo systems,''
\newblock {\em IEEE Transactions on Vehicular Technology}, vol. 70, no. 5, pp. 5115--5120, 2021.

\bibitem{rsprec}
A.~R. Flores, R.~C. de~Lamare, and B.~Clerckx,
\newblock ``Linear precoding and stream combining for rate splitting in multiuser mimo systems,''
\newblock {\em IEEE Communications Letters}, vol. 24, no. 4, pp. 890--894, 2020.

\bibitem{rscf}
A.~R. Flores, R.~C. de~Lamare, and K.~V. Mishra,
\newblock ``Clustered cell-free multi-user multiple-antenna systems with rate-splitting: Precoder design and power allocation,''
\newblock {\em IEEE Transactions on Communications}, vol. 71, no. 10, pp. 5920--5934, 2023.

\bibitem{mbthp}
K.~Zu, R.~C. de~Lamare, and M.~Haardt,
\newblock ``Multi-branch tomlinson-harashima precoding design for mu-mimo systems: Theory and algorithms,''
\newblock {\em IEEE Transactions on Communications}, vol. 62, no. 3, pp. 939--951, 2014.

\bibitem{rmbthp}
L.~Zhang, Y.~Cai, R.~C. de~Lamare, and M.~Zhao,
\newblock ``Robust multibranch tomlinson–harashima precoding design in amplify-and-forward mimo relay systems,''
\newblock {\em IEEE Transactions on Communications}, vol. 62, no. 10, pp. 3476--3490, 2014.

\bibitem{rsthp}
A.~R. Flores, R.~C. De~Lamare, and B.~Clerckx,
\newblock ``Tomlinson-harashima precoded rate-splitting with stream combiners for mu-mimo systems,''
\newblock {\em IEEE Transactions on Communications}, vol. 69, no. 6, pp. 3833--3845, 2021.

\bibitem{rrdstap}
X.~Wang, Z.~Yang, J.~Huang, and R.~C. de~Lamare,
\newblock ``Robust two-stage reduced-dimension sparsity-aware stap for airborne radar with coprime arrays,''
\newblock {\em IEEE Transactions on Signal Processing}, vol. 68, pp. 81--96, 2020.

\bibitem{rcnn}
S.~Mohammadzadeh, V.~H. Nascimento, R.~C. de~Lamare, and N.~Hajarolasvadi,
\newblock ``Robust beamforming based on complex-valued convolutional neural networks for sensor arrays,''
\newblock {\em IEEE Signal Processing Letters}, vol. 29, pp. 2108--2112, 2022.

\bibitem{rapa}
A.~R. Flores and R.~C. de~Lamare,
\newblock ``Robust and adaptive power allocation techniques for rate splitting based mu-mimo systems,''
\newblock {\em IEEE Transactions on Communications}, vol. 70, no. 7, pp. 4656--4670, 2022.

\bibitem{rdiff}
T.~Yu, R.~C. de~Lamare, and Y.~Yu,
\newblock ``Robust resilient diffusion over multi-task networks against byzantine attacks: Design, analysis and applications,''
\newblock {\em IEEE Transactions on Signal Processing}, vol. 70, pp. 2826--2841, 2022.

\bibitem{zcprec}
D.~M.~V. Melo, L.~T.~N. Landau, R.~C. de~Lamare, P.~F. Neuhaus, and G.~P. Fettweis,
\newblock ``Zero-crossing precoding techniques for channels with 1-bit temporal oversampling adcs,''
\newblock {\em IEEE Transactions on Wireless Communications}, vol. 22, no. 8, pp. 5321--5336, 2023.

\bibitem{rrs}
A.~R. Flores and R.~C. de~Lamare,
\newblock ``Robust rate-splitting-based precoding for cell-free mu-mimo systems,''
\newblock {\em IEEE Communications Letters}, vol. 29, no. 6, pp. 1230--1234, 2025.

\bibitem{mashdour2025robust}
S.~Mashdour, A.~R. Flores, S.~Salehi, R.~C. de~Lamare, A.~Schmeink, and P.~B. Da~Silva,
\newblock ``Robust resource allocation in cell-free massive {MIMO} systems,''
\newblock {\em IEEE Transactions on Communications}, 2025.

\bibitem{palhares2021robust}
V.~M.~T. Palhares, A.~R. Flores, and R.~C. De~Lamare,
\newblock ``Robust {MMSE} precoding and power allocation for cell-free massive {MIMO} systems,''
\newblock {\em IEEE Transactions on Vehicular Technology}, vol. 70, no. 5, pp. 5115--5120, 2021.

\bibitem{mashdour2022enhanced}
S.~Mashdour, R.~C. de~Lamare, and Joao~PSH Lima,
\newblock ``Enhanced subset greedy multiuser scheduling in clustered cell-free massive {MIMO} systems,''
\newblock {\em IEEE Communications Letters}, vol. 27, no. 2, pp. 610--614, 2022.

\bibitem{mishra2022rate}
A.~Mishra, Y.~Mao, L.~Sanguinetti, and B.~Clerckx,
\newblock ``Rate-splitting assisted massive machine-type communications in cell-free massive {MIMO},''
\newblock {\em IEEE Communications Letters}, vol. 26, no. 6, pp. 1358--1362, 2022.

\bibitem{hjorungnes2007complex}
A.~Hjorungnes and D.~Gesbert,
\newblock ``Complex-valued matrix differentiation: Techniques and key results,''
\newblock {\em IEEE Transactions on Signal Processing}, vol. 55, no. 6, pp. 2740--2746, 2007.

\end{thebibliography}

\end{document}